\documentclass[%
 reprint,
superscriptaddress,
 amsmath,amssymb,
 aps,
]{revtex4-1}

\usepackage{graphicx}% Include figure files
\usepackage{dcolumn}% Align table columns on decimal point
\usepackage{bm}% bold math
\usepackage{hyperref}% add hypertext capabilities
\usepackage{float}
\begin{document}

\preprint{APS/123-QED}

\title{Testing The Weak Cosmic Censorship Conjecture in Short Haired Black Holes}

\author{Min Zhao}

\author{Meirong Tang}

\author{Zhaoyi Xu}
\homepage{zyxu@gzu.edu.cn(Corresponding author)}

\collaboration{College of Physics, Guizhou University, Guiyang, 550025, China}

\begin{abstract}

The Weak Cosmic Censorship Conjecture is a hypothesis regarding the properties of event horizons and singularities during the formation of black holes, stating that singularities are always encompassed by the event horizons(TEH) of black holes, thus preventing naked singularities from affecting the causal structure of spacetime. In this paper, we explore the Weak Cosmic Censorship Conjecture in the context of rotating hairy black holes, aiming to understand the impact of hairiness on the conjecture for Kerr black holes.We investigate whether TEH of rotating hairy black holes can be disrupted by incoming test particles and scalar fields.When test particles and scalar fields incident on the rotating hairy black hole are found.In extreme cases, when considering a second order approximation, if the parameter $\kappa$ (The parameter here is a function related to hair strength $\kappa(Q_m^{2k})$.)falls within the range of $0<\kappa<\sqrt{1/3}$, TEH of a hairy black hole can be disrupted. Conversely, in the range of $\sqrt{1/3}<\kappa<1$, TEH of a hairy black hole cannot be disrupted.When considering the second order approximation in near extreme cases, the parameter $\kappa$, within the range of $0<\kappa<1$, can lead to the disruption of TEH in this spacetime. When an incident scalar field is present, in near extreme conditions, TEH of a rotating short hair black hole cannot be disrupted. Therefore, the value of the parameter $\kappa$ reveals the connection between rotating short hair black holes and the weak cosmic censorship conjecture, indicating that the presence of short hair significantly affects TEH of black holes. This will aid in further understanding the nature of rotating short hair black holes.\\

\end{abstract}

\maketitle

\section{INTRODUCTION}

Black holes are celestial bodies predicted by the theory of general relativity, and the concept of black holes was first proposed by scientists such as John Michell and Pierre-Simon Laplace in the early 18th century \cite{Michell:1784xqa,Laplace_2009}. The proposal of the black hole concept has been indirectly confirmed by physicists through various observational means, such as gravitational lensing effects, the motion of star orbits, the detection of X-rays, and gravitational waves, all of which provide solid support for the real existence of black holes \cite{Armstrong:2006zz,Ghez:2000ay,Bachlechner:2017hsj,LIGOScientific:2016aoc}. Black holes are composed of extremely dense matter, among which the most notable feature is the singularity within a black hole. Regarding the singularity of black holes, there are two scenarios: one is the gravitational collapse leading to a naked singularity, and the other is that the end product of gravitational collapse is a black hole, meaning the singularity produced is enveloped by the event horizon(TEH) of the black hole, thereby not producing a naked singularity. Naked and non naked singularities represent the two possibilities of whether the singularity can be covered by TEH. A naked singularity means that the singularity is not covered by TEH and can be directly exposed to external observation; whereas a non naked singularity means the singularity is completely surrounded by TEH, not affecting the completeness of spacetime causality. This issue actually relates to the weak cosmic censorship conjecture, which is a conjecture about the properties of black holes. The weak cosmic censorship conjecture suggests that all singularities in the universe should be enveloped by TEH of black holes, preventing them from affecting the causal laws of spacetime.

In the previous section, we discussed two important concepts: singularities and the Weak Cosmic Censorship Conjecture.There are some connections between them. On one hand, an important result of General Relativity is the Singularity Theorems, proposed by physicists Penrose and Hawking\cite{Penrose:1964wq,Hawking:1970zqf}, which suggest that matter collapse inevitably leads to singularities, thus causing the breakdown of physical laws. On the other hand, to ensure that physical laws remain unaffected by singularities, Penrose proposed the Weak Cosmic Censorship Conjecture in 1979 \cite{Penrose:1969pc,Carroll_2019}. Specifically, the idea behind the Weak Cosmic Censorship Conjecture is that the appearance of singularities is always accompanied by event horizons(EH), thereby preventing singularities from being exposed to the universe. To test this conjecture, physicists have used various methods such as the collapse evolution of matter fields\cite{East:2019bwu,Song:2020onc,Goswami:2007na}, collisions of supermassive black holes\cite{Andrade:2019edf,Andrade:2018yqu,Brill:1993tm,Choptuik:2003as}, and numerical evolutions of black holes \cite{Ahmed:2022dpu,Nie:2021rhz}. Here, we will employ the thought experiment proposed by Wald to test the weak cosmic censorship conjecture, by injecting test particles into both extreme and near extreme black holes \cite{Wald:1974hkz}, as well as the method proposed by Semiz and others, which involves injecting a scalar field into extreme and near extreme black holes to test the weak cosmic censorship conjecture \cite{Semiz:2005gs}. Research on this conjecture has led to different schools of thought \cite{Gao:2012ca,Gwak:2015fsa,Ghosh:2019dzq,Siahaan:2015ljs,Shaymatov:2022ako,Ying:2020bch,Chirco:2010rq,deFelice:2001wj,He:2019kws,Chen:2019pdj,Matsas:2007bj,Richartz:2008xm,Rocha:2014jma,Duztas:2016xfg,Gwak:2017kkt,Meng:2024els,Zhao:2024lts,Tang:2023sig,Meng:2023vkg,Zhao:2023vxq}, with some studies suggesting that the Weak Cosmic Censorship Conjecture could be violated, exposing the singularities of black holes, while other studies argue that it is not violated, thus ensuring that black holes' event horizons encompass them. Therefore, research on the Weak Cosmic Censorship Conjecture is continuously evolving and changing.

On the other hand, through the Einstein field equations, exact solutions for black holes can be obtained, allowing physicists to gain a deeper understanding of the nature of black holes. Typically, the exact solutions for black holes are divided into spherically symmetric and rotating cases. Physicists know through research that gravitational collapse can form black holes, and mass alone can describe the properties of spherically symmetric black holes. However, there is another type of black hole, namely the rotating Kerr black hole, which is described by its mass, angular momentum, and charge, according to the no hair theorem \cite{Israel:1967wq,Israel:1967za,Carter:1971zc,Hawking:1971vc,Robinson:1975bv,Mazur:1982db,Bekenstein:1995un,Broderick:2013rlq,Isi:2019aib,Wang:2021elt,Gurlebeck:2015xpa,Herdeiro:2015waa}. Among the many "hairs" of black holes, scalar hair, as the most important one, affects the properties of black holes. Tang and others have used the NJ (Newman-Janis) method to extend the background metric of spherically symmetric short haired black holes to the background metric of rotating shor haired black holes \cite{Tang:2022uwi}. In this work, by injecting test particles and scalar fields into the rotating short haired black holes calculated by Tang and others, the influence of short hair on the Weak Cosmic Censorship Conjecture for Kerr black holes is explored.

The structure of this article is as follows: the first section is the introduction. The second section introduces the exact solutions of rotating short haired black holes. The third section discusses the injection of test particles into rotating short haired black holes. The fourth section covers the injection of scalar fields into rotating short haired black holes. The fifth section provides a summary of the article. This article adopts natural units where $c = G = 1$.

\section{Short haired black holes under rotation}

\subsection{Precise solution of short haired black holes in the case of rotation}

The no hair theorem is a key feature of classical black holes, and the quantum effects of black hole EH become particularly significant due to the extreme properties of trapped ergospheric horizons \cite{Hawking:2016msc,Herdeiro:2014goa,Herdeiro:2015gia,Dvali:2012rt,Coleman:1991ku,PhysRevLett.67.1975,Bousso:2017dny}. However, the non trivial matter fields in black hole spacetimes may lead to violations of the no hair theorem. Among them, scalar hair acts as the most significant hair of black holes, affecting the spacetime structure of black holes.

E. Contreras and others have utilized the gravitational decoupling method to extend these spherically symmetric black holes with scalar hair to the rotating case, discussing the fundamental physical properties of such spacetimes \cite{Contreras:2021yxe}. This is based on the spherically symmetric hairy background metrics obtained by J. Ovalle and others \cite{Ovalle:2020kpd}. Using the Newman-Janis (NJ) algorithm, solutions for rotating short haired black holes were obtained, meaning that the NJ method can extend spherically symmetric spacetimes to rotating spacetimes through complex transformations. Therefore, within the NJ algorithm \cite{Newman:1965tw,Azreg-Ainou:2014pra,Azreg-Ainou:2014nra}, Tang and others \cite{Tang:2022uwi} obtained the exact solutions for short haired black holes in the rotating case, as follows 

\begin{widetext}
\begin{equation}\label{z1}
ds^2=-\bigg[1-\frac{2Mr-\frac{Q_m^{2k}}{r^{2k-2}}}{\rho^2}\bigg]dt^2+\frac{\rho^2}{\Delta}dr^2-\frac{2a \sin^2\theta(2Mr-\frac{Q_m^{2k}}{r^{2k-2}})}{\rho^2}dtd\phi+\rho^2d\theta^2+\frac{\Sigma\sin^2\theta}{\rho^2}d\phi^2,
\end{equation}
\end{widetext}

here
\begin{equation}
\rho^2=r^2+a^2\cos^2\theta,
\end{equation}

\begin{equation}\label{z2}
\Delta=r^2-2Mr+\frac{Q_m^{2k}}{r^{2k-2}}+a^2,
\end{equation}

\begin{equation}
\Sigma=(r^2+a^2)^2-a^2\Delta(r)\sin^2\theta,
\end{equation}

The mass of the black hole is denoted by the variable M in the given equation, $Q_m$ denotes the strength parameter of the hair, where for $Q_m$ $\neq $ 0 and $k>1$, it represents a short haired black hole under rotation. Here, $a$ represents the spin parameter of the short haired black hole, namely $a=J/M$, where $J$ denotes the angular momentum of the black hole. When $Q_m$=0, the metric (\ref{z1}) representing the short haired black hole degenerates to a Kerr black hole.

\subsection{Event horizon and angular velocity of a short haired black hole}

TEH of this short haired black hole can be expressed in a coupled manner, given by $g^{rr}$=$\Delta$=0. This can be obtained through equation (\ref{z2}).

\begin{equation}\label{z3}
\Delta=r^2-2Mr+\frac{Q_m^{2k}}{r^{2k-2}}+a^2=0,
\end{equation}

By rearranging equation (\ref{z3}), we obtain a coupled equation for TEH
\begin{equation}\label{z4}
r_h=M\pm M\sqrt{1-\frac{\frac{Q_m^{2k}}{r_h^{2k-2}}+a^2}{M^2}},
\end{equation}

For convenience of analysis, let the coupling term be $\beta=\frac{Q_m^{2k}}{r_h^{2k-2}}$. Then equation (\ref{z4}) can be rewritten as follows

\begin{equation}\label{z5}
r_h=M\pm M\sqrt{1-\frac{\beta+a^2}{M^2}},
\end{equation}

Where $r_h$ represents TEH of the rotating short haired black hole. From equation (\ref{z5}), we know that when $\beta+a^2<M^2$ is satisfied, it indicates the existence of a black hole spacetime. However, in the other case when $\beta+a^2>M^2$ is satisfied, the background metric (\ref{z1}) no longer describes a black hole. This is the point of interest for us because if, after scattering test particles or scalar fields, the metric (\ref{z1}) transitions from describing a black hole to a non black hole spacetime, then it can be said that the metric (\ref{z1}) has undergone super spinning. In other words, at this point, there is a possibility of violating The Weak Cosmic Censorship Conjecture under our conditions.

The formula provided below expresses the area of TEH for this particular black hole:
\begin{equation}
A=\iint \sqrt{g_{\theta \theta}g_{\phi \phi}}d\theta d\phi=4\pi(r_h^2+a^2),
\end{equation}

At TEH of this black hole, the angular velocity can be determined using the following formula:
\begin{equation}\label{z10}
\Omega_h=-\frac{g_{03}}{g_{33}}=\frac{a}{r_h^2+a^2}.
\end{equation}

\section{Incident test particles on rotating short haired black holes}

This chapter, our main focus is on incident test particles in rotating short haired black holes, aiming to discuss the possibility of disrupting the black hole EH in extreme and near extreme black hole scenarios. Through equation (\ref{z5}), we can calculate TEH of rotating short haired black holes. In this equation, when $\beta+a^2\leq M^2$ satisfies certain conditions, the black hole EH exists. However, when it exceeds a certain threshold denoted by $\beta+a^2>M^2$, it implies the absence of an EH in this spacetime, which is precisely the issue we are primarily discussing. Therefore, in this article, we primarily delve into this problem.

The movement of particles in the spacetime of rotating short haired black holes can be described using the geodesic equation
\begin{equation}
\frac{d^2x^{\mu}}{d\tau^2}+\Gamma_{\alpha \beta}^{\mu}\frac{dx^{\alpha}}{d\tau}\frac{dx^{\beta}}{d\tau}=0,
\end{equation}

The Lagrangian is given by:
\begin{equation}
L=\frac{1}{2}mg_{\mu\nu}\frac{dx^{\mu}}{d\tau}\frac{dx^{\nu}}{d\tau}=\frac{1}{2}mg_{\mu\nu}\dot{x}^{\mu}\dot{x}^{\nu},
\end{equation}

If a test particle moves at a slow pace along the equatorial plane with $\theta=\frac{\pi}{2}$, it does not exhibit any motion in the $\theta$direction,  resulting in $d\theta/d\tau$=0, which implies that the momentum $P_\theta$ is zero.

\begin{equation}
P_{\theta}=\frac{\partial L}{\partial \dot{\theta}}=mg_{22}\dot{\theta}=0,
\end{equation}

From the equations of motion of the test particle, the angular momentum $\delta J$ and energy $\delta E$ can be expressed as the components of the test particle in the $\phi$ and $t$ directions, respectively. Their expressions are as follows

\begin{equation}
\delta J=P_{\phi}=\frac{\partial L}{\partial \phi}=mg_{3\nu}\dot{x}^{\nu},
\end{equation}

\begin{equation}
\delta E=-P_t=-\frac{\partial L}{\partial \dot{t}}=-mg_{0\nu}\dot{x}^{\nu}.
\end{equation}

When a test particle enters the interior of TEH of a rotating short haired black hole, the energy and angular momentum of the short haired black hole change. At this point, the changed energy and angular momentum are as follows
\begin{equation}\label{z12}
M\to M^{'}=M+\delta E,
\end{equation}

\begin{equation}\label{z13}
J\to J^{'}=J+\delta J,
\end{equation}

When the test particle moves outside TEH of the short haired black hole, its four velocity is given as follows
\begin{equation}
U^{\mu}U_{\mu}=\frac{dx^{\mu}}{d\tau}\frac{dx_{\mu}}{d\tau}=g_{\mu\nu}\frac{dx^{\mu}}{d\tau}\frac{dx^{\nu}}{d\tau}=\frac{1}{m^2}g^{\mu\nu}P_{\mu}P_{\nu}=-1,
\end{equation}

Substituting the expressions for angular momentum and energy into the above equation and rearranging, we obtain the following expression
\begin{equation}
g^{00}\delta E^2-2g^{03}\delta E\delta J+g^{11}P_r^2+g^{33}\delta J^2=-m^2,
\end{equation}

The equation $\delta E$ above can be calculated as
\begin{equation}
\delta E=\frac{g^{03}}{g^{00}}\delta J\pm \frac{1}{g^{00}}\sqrt{(g^{03})^2\delta J^2-g^{00}(g^{33}\delta J^2+g^{11}P_r^2+m^2)},
\end{equation}

As the test particle travels from infinity towards TEH, its trajectory must be timelike and oriented towards the future. Therefore, in this case, we need to satisfy the following conditions
\begin{equation}\label{z6}
\frac{dt}{d\tau}>0,
\end{equation}

For the two equations of $\delta J$ and $\delta E$, rearranging yields
\begin{equation}\label{z7}
\dot{t}=\frac{dt}{d\tau}=-\frac{g_{33}\delta E+g_{03}\delta J}{g_{00}g_{33}-g_{03}^2},
\end{equation}

The energy is obtained from equations (\ref{z6}) and (\ref{z7})
\begin{equation}\label{z8}
\delta E>-\frac{g_{03}}{g_{33}}\delta J,
\end{equation}

At this point, we need to satisfy the condition of equation (\ref{z8}), which means the energy of the test particle takes a negative sign, that is
\begin{equation}
\delta E=\frac{g^{03}}{g^{00}}\delta J-\frac{1}{g^{00}}\sqrt{(g^{03})^2\delta J^2-g^{00}(g^{33}\delta J^2+g^{11}P_r^2+m^2)}.
\end{equation}

In studying the constraints on energy and angular momentum, we conclude that to ensure the test particle can accurately enter TEH, its energy and angular momentum must meet specific requirements, that is
\begin{equation}\label{z9}
\delta J<-\lim\limits_{r\to r_h}\frac{g_{33}}{g_{03}}\delta E,
\end{equation}

By combining equations (\ref{z10}) and (\ref{z9}), the solution obtained is
\begin{equation}\label{z11}
\delta J<-\lim\limits_{r\to r_h}\frac{g_{33}}{g_{03}}\delta E=\frac{\delta E}{\Omega_h}=\frac{r^2+a^2}{a}\delta E,
\end{equation}

When the angular momentum is too large, it is impossible for the test particle to fall into the black hole. Therefore, for a test particle to be captured by a rotating short haired black hole, there must be an upper limit to the test particle's angular momentum, denoted as $\delta J_{max}$. Thus, the upper limit of the test particle's angular momentum $\delta J$, given by equation (\ref{z11}), is

\begin{equation}\label{z17}
\delta J_{max}<\frac{\delta E}{\Omega_h}=\frac{r_h^2+a^2}{a}\delta E.
\end{equation}

From equation (\ref{z5}), we know that TEH of the rotating short haired black hole disappears only when $\beta+a^2>M^2$. Therefore, we can arrange this condition as follows

\begin{equation}\label{z19}
a>M\sqrt{1-\frac{\beta}{M^2}}=M\kappa,
\end{equation}

that is
\begin{equation}
J>M^2\kappa,
\end{equation}
Here, the parameter $\kappa=\sqrt{1-\frac{\beta}{M^2}}$ is a function of the hair strength $\kappa(Q_m^{2k})$, where $\beta=\frac{Q_m^{2k}}{r_h^{2k-2}}$, and the parameter $\kappa$ is the Greek letter, while the k in the hair strength $Q_m^{2k}$ is the letter k.

Therefore, for a test particle entering a rotating short haired black hole to disrupt TEH of the black hole, the following condition is also required
\begin{equation}\label{z15}
J^{'}>\kappa^{'}M^{{'}2},
\end{equation}

Since our $\kappa^{'}$ is $\sqrt{1-\frac{\beta}{(M+\delta E)^2}}$, it means that at this moment $\kappa^{'}$ contains a mass coupling term. Here, since $\delta E\ll M$, expanding the coupling term yields

\begin{equation}\label{z14}
\kappa^{'}\approx \kappa+\frac{\beta}{\kappa M^3}\delta E-\bigg(\frac{\beta^2}{2\kappa^3M^6}+\frac{3\beta}{2\kappa M^4}\bigg)\delta E^2,
\end{equation}

Substituting equations (\ref{z12}), (\ref{z13}), and (\ref{z14}) into equation (\ref{z15}) yields
\begin{widetext}
\begin{equation}
J+\delta J>\bigg(\kappa+\frac{\beta}{\kappa M^3}\delta E-\bigg(\frac{\beta^2}{2\kappa^3M^6}+\frac{3\beta}{2\kappa M^4}\bigg)\delta E^2\bigg)(M^2+2M\delta E+\delta E^2),
\end{equation}
\end{widetext}

After rearranging, we obtain
\begin{widetext}
\begin{equation}\label{z16}
\delta J>(M^2\kappa-J)+\bigg(2\kappa M+\frac{\beta}{\kappa M}\bigg)\delta E+\bigg(\kappa+\frac{2\beta}{\kappa M^2}-\frac{\beta^2}{2\kappa^3M^4}-\frac{3\beta}{2\kappa M^2}\bigg)\delta E^2,
\end{equation}
\end{widetext}

In equation (\ref{z16}), we obtain a lower limit for the test particle to disrupt TEH of the short haired black hole in the rotating case. When we do not consider the effects of higher order perturbations, a lower limit for the angular momentum is obtained
\begin{equation}\label{z18}
\delta J_{min}>(M^2\kappa-J)+\bigg(2\kappa M+\frac{\beta^2}{\kappa M}\bigg)\delta E,
\end{equation}

Through the above derivation, we know that when the chosen test particle simultaneously satisfies the conditions of equations (\ref{z17}) and (\ref{z18}), TEH of the short haired black hole in the rotating case can potentially be disrupted, which minimum limit would reveal the internal configuration of the black hole.

Moving forward, we will delve into the scenarios of extreme and near extreme cases.

1$\star$ In the extreme case, where $\frac{\beta+a^2}{M^2}=1$, TEH of the rotating short haired black hole is as follows
\begin{equation}\label{z20}
r_h=M,
\end{equation}

In the case of first order approximation, to disrupt TEH of the rotating short haired black hole, the following two conditions need to be satisfied
\begin{equation}\label{z21}
\delta J_{max}<\frac{\delta E}{\Omega_h}=\frac{r_h^2+a^2}{a}\delta E,
\end{equation}

\begin{equation}\label{z22}
\delta J_{min}>\bigg(2\kappa M+\frac{\beta}{\kappa M}\bigg)\delta E,
\end{equation}

By combining equations (\ref{z19}), (\ref{z20}), and (\ref{z21}), it can be calculated that
\begin{widetext}
\begin{equation}\label{z23}
\delta J_{max}<\frac{r_h^2+a^2}{a}\delta E=\frac{M^2\kappa^2+M^2\kappa^2+\beta}{M\kappa}\delta E=\frac{2M^2\kappa^2+\beta}{M\kappa}\delta E=\bigg(2M\kappa+\frac{\beta}{M\kappa}\bigg)\delta E,
\end{equation}
\end{widetext}

Analyzing equations (\ref{z22}) and (\ref{z23}), in the case of a first order approximation, it can be intuitively obtained that
\begin{equation}
\delta J_{max}-\delta J_{min}=\bigg(2M\kappa +\frac{\beta}{M\kappa }\bigg)\delta E-\bigg(2M\kappa +\frac{\beta}{M\kappa}\bigg)\delta E=0,
\end{equation}

So in the first order case, we get the same maximum and minimum values for the test particle, and then we get the case that TEH of the black hole is not destroyed only in the first order case. This forces us to consider second order minuses, which, when taken into account, can be obtained according to equations (\ref{z16}) and (\ref{z23}).

\begin{widetext}
\begin{equation}
\delta J_{max}-\delta J_{min}=\bigg(2M\kappa +\frac{\beta}{M\kappa }\bigg)\delta E-\bigg(2M\kappa +\frac{\beta}{M\kappa }\bigg)\delta E-\bigg(\kappa+\frac{2\beta}{\kappa M^2}-\frac{\beta^2}{2\kappa^3M^4}-\frac{3\beta}{2\kappa M^2}\bigg)\delta E^2=-\bigg(\kappa+\frac{\beta}{2\kappa M^2}-\frac{\beta^2}{2\kappa^3M^4}\bigg)\delta E^2,
\end{equation}
\end{widetext}

In this equation, we notice that a term $M^2\kappa-J$ is missing, which is because in the extreme case $M^2\kappa=J$.Converting $\kappa=\sqrt{1-\frac{\beta}{M^2}}$ yields
\begin{equation}
\beta=M^2-M^2\kappa^2,
\end{equation}

\begin{widetext}
\begin{eqnarray}\label{z24}
\delta J_{max}-\delta J_{min}&&=-\bigg(\kappa+\frac{\beta}{2\kappa M^2}-\frac{\beta^2}{2\kappa^3 M^4}\bigg)\delta E^2=-\bigg(\kappa+\frac{M^2-M^2\kappa^2}{2\kappa M^2}-\frac{M^4+M^4\kappa^4-2M^4\kappa^2}{2\kappa^3 M^4}\bigg)\delta E^2 \nonumber\\
&&=-\bigg(\kappa +\frac{1}{2\kappa}-\frac{1}{2}\kappa-\frac{1}{2\kappa^3}-\frac{1}{2}\kappa+\frac{1}{\kappa}\bigg)\delta E^2=-\bigg(\frac{3}{2\kappa}-\frac{1}{2\kappa^3}\bigg)\delta E^2=-\frac{1}{2\kappa}(3-\frac{1}{\kappa^2})\delta E^2,
\end{eqnarray}
\end{widetext}

\begin{figure}[H]
   \centering
     \includegraphics[width=\columnwidth]{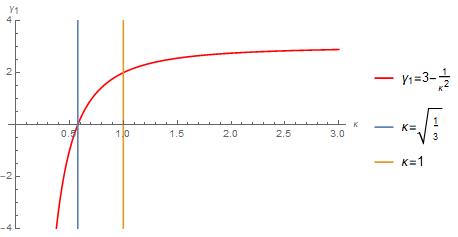}% Here is how to import EPS art
     \caption{}
     \label{fig1}
\end{figure}

In equation (\ref{z24}), let $\gamma_1<3-\frac{1}{\kappa^2}$, when $\gamma_1<0$, then $\delta J_{max}-\delta J_{min}>0$, which means that in the extreme case, TEH of the rotating short haired black hole can be disrupted. When $\kappa$ satisfies $3-\frac{1}{\kappa^2}<0$, TEH of the black hole can be disturbed, from which $-\sqrt{\frac{1}{3}}<\kappa<\sqrt{\frac{1}{3}}$ is obtained. Since the coupling term $\beta<\frac{Q_m^{2k}}{r_h^{2k-2}}$ is positive, it is known from $\kappa=\sqrt{1-\frac{\beta}{M^2}}$ that if this equation holds, then $\kappa$ also needs to satisfy the condition $0<\kappa<1$. Therefore, through this analysis, we know that when the incident test particle disrupts TEH of the rotating short haired black hole in the extreme case, the range of $\kappa$ is $0<\kappa<\sqrt{\frac{1}{3}}$. This range is clearly visible in Figure (\ref{fig1}).

Figure (\ref{fig1}): According to the analysis above, the range of $\kappa$ is between $0<\kappa<1$. One scenario is when $0<\kappa<\sqrt{\frac{1}{3}}$, which corresponds to the range on the left side of the blue line in Figure (\ref{fig1}) where TEH of the black hole can be disrupted. Another scenario is between the blue and yellow lines, that is, in the range of $\sqrt{\frac{1}{3}}<\kappa<1$, where $\gamma_1$ is positive. This indicates that $\delta J_{max}-\delta J_{min}<0$, in other words, within this range, TEH of the rotating short haired black hole cannot be disrupted.

2$\star$ In the near extreme case, which is when $a\sim \kappa M$, the condition for a test particle to disrupt TEH of the rotating short haired black hole is as follows
\begin{equation}\label{z26}
\delta J_{max}<\frac{r_h^2+a^2}{a}\delta E,
\end{equation}

\begin{equation}\label{z27}
\delta J_{min}>\bigg(2\kappa M+\frac{\beta}{\kappa M}\bigg)\delta E+(\kappa M^2-J),
\end{equation}

The closeness of $a\sim \kappa M$ can be described by a dimensionless small quantity $\epsilon$
\begin{equation}\label{z25}
\frac{\beta+a^2}{M^2}=1-\epsilon^2,
\end{equation}

Parameter $\epsilon \to 0$, that is $\epsilon \ll 1$, when $\epsilon=0$, equation (\ref{z25}) represents the extreme case. From equations (\ref{z26}) and (\ref{z27}), it can be concluded that in the near extreme situation, TEH of the short haired black hole spacetime is disrupted. Here, ($\kappa M^2-J$) is a second order small quantity, thus obtained
\begin{equation}\label{z28}
\frac{1}{\Omega_h}-2\kappa M-\frac{\beta}{\kappa M}>0,
\end{equation}

This indicates that under the condition satisfied, TEH of the short haired black hole in rotation can be disrupted in the near extreme situation.

Since $\epsilon \ll 1$, some series expansions are made
\begin{equation}\label{z29}
r_h=M(1+\epsilon),
\end{equation}

\begin{equation}\label{z30}
a=M\bigg(\kappa-\frac{\epsilon^2}{2\kappa}+o(\epsilon^4)\bigg),
\end{equation}

\begin{equation}
a^2=M^2-M^2\epsilon^2-\beta,
\end{equation}

Combining equations (\ref{z25}), (\ref{z28}), (\ref{z29}), and (\ref{z30}), the calculation is obtained
\begin{equation}\label{z31}
\frac{1}{\Omega_h}-2M\kappa-\frac{\beta}{\kappa M}=\frac{2M^2\epsilon+\bigg(M^2+\frac{\beta}{2\kappa^2}\bigg)\epsilon^2-o(\epsilon^4)}{a}.
\end{equation}

In equation (\ref{z31}), $\epsilon$ is a first order small quantity, and $\delta \epsilon$ is also a first order small quantity. Here, the product of $\epsilon$ and $\delta \epsilon$ is a second order small quantity, that is, $\epsilon \cdot \delta \epsilon \sim o(\delta E^2)$. This leads to the entire equation presenting a second order situation, which compels us to consider the case of $\delta E^2$.

It can be obtained from the spin parameter $a=\frac{J}{M}$
\begin{equation}
M^2\kappa-J=\frac{M^2\epsilon^2}{2\kappa}-o(\epsilon^4).
\end{equation}

It is obtained from equations (\ref{z16}) and (\ref{z31})
\begin{equation}
\delta J_{max}-\delta J_{min}=\bigg(\frac{2M^2}{a}-\bigg(\frac{2}{2\kappa}-\frac{1}{2\kappa^3}\bigg)-\frac{M^2}{2\kappa}\bigg)\delta E^2+o(\delta E^3),
\end{equation}

let

\begin{equation}
\gamma=\frac{2M^2}{a}-\bigg(\frac{3}{2\kappa}-\frac{1}{2\kappa^3}\bigg)-\frac{M^2}{2\kappa}.
\end{equation}

\begin{figure}[H]
   \centering
     \includegraphics[width=\columnwidth]{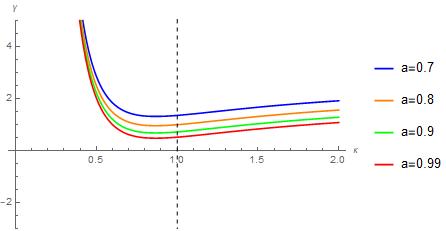}% Here is how to import EPS art
     \caption{}
     \label{fig2}
\end{figure}

In the analysis of the extreme case, we know that the range of $\kappa$ is $0<\kappa<1$, as indicated by the left side of the dashed line in Figure (\ref{fig2}). From Figure (\ref{fig2}), it is known that within this range, $\gamma$ is always positive. This means that in the near extreme situation, within the range of $0<\kappa<1$, TEH of the rotating short haired black hole can be disrupted.

\section{Incident scalar field in a rotating short haired black hole}

In the third section, the weak cosmic censorship conjecture was tested using the method of test particles. This section primarily utilizes the method proposed by Semiz and others \cite{Semiz:2005gs}, introducing a massive scalar field into the spacetime of a rotating short haired black hole to discuss whether TEH of the black hole can be disrupted in the cases of extreme and near extreme black holes.

\subsection{Scattering of a massive scalar field}

In the scalar field, the Klein Gordon equation under curved spacetime is as follows
\begin{equation}
\frac{1}{\sqrt{-g}}\partial_{\mu}(\sqrt{-g}g^{\mu\nu}\partial_{\nu}\psi)-\mu^2\psi=0.
\end{equation}

Based on the metric given by equation (\ref{z1}), its determinant can be calculated
\begin{equation}
g=-\rho^4\sin^2\theta,
\end{equation}

the inverse metric tensor of the metric (\ref{z1}) is given by the following equation
\begin{equation}
g^{\mu\nu}=\frac{\Delta^{\mu\nu}}{g}.
\end{equation}
Let $\eta=2Mr-\frac{Q_m^{2k}}{r^{2k-2}}$, substituting the metric (\ref{z1}) into the above expression yields the following equation:

\begin{widetext}
\begin{eqnarray}\label{z33}
&&-\frac{(r^2+a^2)^2-a^2\Delta \sin^2\theta}{\Delta \rho^2}\frac{\partial^2\psi}{\partial t^2}-\frac{2a\eta}{\Delta \rho^2}\frac{\partial^2\psi}{\partial t\partial \phi}+\frac{1}{\rho^2}\frac{\partial}{\partial r}\bigg(\Delta \frac{\partial \psi}{\partial r}\bigg)+\frac{1}{\rho^2\sin \theta}\frac{\partial}{\partial \theta}\bigg(\sin \theta \frac{\partial \psi}{\partial \theta}\bigg)\nonumber\\
&&+\frac{\Delta-a^2\sin^2\theta}{\Delta \rho^2 \sin^2\theta}\frac{\partial^2\psi}{\partial \phi^2}-\mu^2\psi=0.
\end{eqnarray}
\end{widetext}

The form of the solution for the scalar field $\psi$ in the above equation is as follows:
\begin{equation}\label{z32}
\psi(t,r,\theta,\phi)=e^{-i\omega t}R(r)S_{lm}(\theta)e^{im\phi},
\end{equation}

Where $S_{lm}(\theta)$ are the angular spherical functions, and $l, m$ are constants of the angular separation variables, whose values are positive integers. Substituting equation (\ref{z32}) into the scalar field equation (\ref{z33}) yields the angular and radial equations, respectively.
\begin{widetext}
\begin{equation}\label{z34}
\frac{1}{\sin \theta}\frac{d}{d\theta}\bigg(\sin \theta \frac{d S_{lm}(\theta)}{d\theta}\bigg)-\bigg(a^2\omega^2\sin^2\theta+\frac{m^2}{\sin^2\theta}+\mu^2a^2\cos^2\theta-\lambda_{lm}\bigg)S_{lm}=0,
\end{equation}

\begin{equation}\label{z35}
\frac{d}{dr}\bigg(\Delta \frac{dR(r)}{dr}\bigg)+\bigg(\frac{(r^2+a^2)^2}{\Delta}\omega^2-\frac{2a\eta}{\Delta}m\omega+\frac{m^2a^2}{\Delta}-\mu^2r^2-\lambda_{lm}\bigg)S_{lm}=0,
\end{equation}
\end{widetext}

The solution to equation (\ref{z34}) is a spherical function, which has an integral of 1 when calculating the energy flux. Therefore, we will solve the radial equation, introducing tortoise coordinates in this calculation process
\begin{equation}
dr_*=\frac{r^2+a^2}{\Delta}dr,
\end{equation}

By substituting the turtle coordinates into the radial equation (\ref{z35}), we get
\begin{widetext}
\begin{equation}\label{z37}
\frac{\Delta}{(r^2+a^2)^2}\frac{d}{dr}(r^2)\frac{dR(r)}{dr_*}+\frac{d^2R(r)}{dr_*^2}+\bigg[\bigg(\omega-\frac{ma}{r^2+a^2}\bigg)^2+\frac{\Delta 2am\omega}{(r^2+a^2)^2}-\frac{\Delta}{(r^2+a^2)^2}(\mu^2r^2+\lambda_{lm})\bigg]R(r)=0.
\end{equation}
\end{widetext}

Near TEH ($r\cong r_h$), namely
\begin{equation}\label{z36}
\Delta \cong0.
\end{equation}

Substituting equation (\ref{z36}) into (\ref{z37}) yields
\begin{equation}\label{z38}
\frac{d^2R(r)}{dr_*^2}+\bigg(\omega-\frac{ma}{r^2+a^2}\bigg)^2R(r)=0.
\end{equation}

Substituting equation (\ref{z10}) into (\ref{z38}) yields
\begin{equation}\label{z39}
\frac{d^2R(r)}{dr_*^2}+(\omega-m\Omega_h)^2R(r)=0.
\end{equation}

Expressed in exponential form
\begin{equation}\label{z40}
R(r)\sim e^{[\pm i(\omega-m\Omega_h)r_*]}.
\end{equation}
Here, the positive and negative signs correspond to the outgoing and incoming waves, respectively. When we inject a scalar field into a rotating shorthaired black hole, the spacetime of this black hole will absorb the energy of the perturbation field. Therefore, it is more in line with physical reality to take the negative sign in equation (\ref{z40}). At this time, the solution to equation (\ref{z39}) is

\begin{equation}\label{z41}
R(r)=e^{[-i(\omega-m\Omega_h)r_*]}.
\end{equation}

By substituting equation (\ref{z41}) into equation (\ref{z32}), we can obtain the approximate solution to the field equation as
\begin{equation}
\psi(t,r,\theta,\phi)=e^{[-i(\omega-m\Omega_h)r_*]}e^{-i\omega t}S_{lm}(\theta)e^{im\phi}.
\end{equation}

Once this solution is acquired, it enables the calculation of the energy and angular momentum that the spacetime of a short haired black hole absorbs when the scalar field is scattered onto it.

The following equation represents the energy momentum tensor of a scalar field $\psi$ with mass $\mu$
\begin{equation}\label{z42}
T_{\mu\nu}=\partial_{\mu}\psi \partial_{\nu}\psi^*-\frac{1}{2}g_{\mu\nu}(\partial_{\mu}\psi \partial^{\nu}\psi^*+\mu^2\psi \psi^*).
\end{equation}

By substituting the background metric of the short haired black hole from equation (\ref{z1}) into equation (\ref{z42}), the following tensor is obtained
\begin{equation}
T_t^r=\frac{r_h^2+a^2}{\rho^2}\omega(\omega-m\Omega_h)S_{lm}(\theta)e^{im\phi}S^*_{l^{'}m^{'}}(\theta)e^{-im\phi},
\end{equation}

\begin{equation}
T_{\phi}^r=\frac{r_h^2+a^2}{\rho^2}m(\omega-m\Omega_h)S_{lm}(\theta)e^{im\phi}S^*_{l^{'}m^{'}}(\theta)e^{-im\phi}.
\end{equation}

The energy flux of TEH in the spacetime of a short haired black hole
\begin{equation}\label{z43}
\frac{dE}{dt}=\iint  T_t^r\sqrt{-g}d\theta d\phi=\omega(\omega-m\Omega_h)[r_h^2+a^2].
\end{equation}
The angular momentum flux through TEH in the spacetime of a short haired black hole
\begin{equation}\label{z44}
\frac{dJ}{dt}=\iint T_{\phi}^r\sqrt{-g}d\theta d\phi=m(\omega-m\Omega_h)[r_h^2+a^2].
\end{equation}

Between equations (\ref{z43}) and (\ref{z44}), one scenario, as we learn from equation $\omega-m\Omega_h$, occurs when $\omega>m\Omega_h$, leading to $\omega-m\Omega_h$ being positive. This indicates that both angular momentum and energy flux are positive, suggesting that the energy $dE$ and angular momentum $dJ$ can be extracted from the scalar field by the rotating short haired black hole. In another scenario, when $\omega<m\Omega_h$, both angular momentum and energy are negative, indicating that the energy extracted by the rotating short haired black hole in the scalar field is due to black hole superradiance \cite{Brito:2015oca}.
During the time interval $dt$, the equations $\delta E$ and $\delta J$ for the rotating short haired black hole are as follows:
\begin{equation}\label{z46}
dE=\omega(\omega-m\Omega_h)[r_h^2+a^2]dt,
\end{equation}
\begin{equation}\label{z47}
dJ=m(\omega-m\Omega_h)[r_h^2+a^2]dt,
\end{equation}
From these two equations, the energy and angular momentum extracted by the short haired black hole from the scalar field are derived. Using these equations, one can discuss the impact on TEH of the black hole due to the scattering of the scalar field onto the rotating short haired black hole in extreme and near extreme conditions.

\subsection{Incident scalar field in a short haired black hole.}
This section primarily investigates the impact of a scalar field encountering a rotating short haired black hole, exploring whether the scalar field with significant angular momentum can disturb TEH of this spacetime. The process of scalar field scattering is discussed using the concept of differentiation, focusing on the time interval $dt$.

During this process, after the rotating short haired black hole absorbs the energy and angular momentum from the incident scalar field, its mass $M$ and angular momentum $J$ become $M^{'}$ and $J^{'}$, respectively. After obtaining the changed mass and angular momentum, we discuss based on equation $\kappa^{'}M^{'2}-J^{'}$. If $\kappa^{'}M^{'2}\ge J^{'}$, then $\kappa^{'}M^{'2}-J^{'}$ is positive, meaning TEH of the rotating short haired black hole exists. Conversely, TEH of this black hole spacetime is disrupted.

Therefore, for the system formed by equation $\kappa^{'}M^{'2}-J^{'}$, after absorbing energy and angular momentum from the scalar field, its state can become
\begin{widetext}
\begin{equation}\label{z45}
\kappa^{'}M^{'2}-J^{'}=(M^2\kappa-J)+\bigg(2M\kappa+\frac{\beta}{M\kappa}\bigg)\delta E+\bigg(\kappa+\frac{2\beta}{\kappa M^2}-\frac{\beta^2}{2M^4\kappa^3}-\frac{3\beta}{2\kappa M^2}\bigg)\delta E^2+o\delta E^3-\delta J,
\end{equation}
\end{widetext}

When higher vorder perturbations are not considered, equation (\ref{z45}) becomes
\begin{equation}\label{z48}
\kappa^{'}M^{'2}-J^{'}=(M^2\kappa-J)+\bigg(2M\kappa+\frac{\beta}{M\kappa}\bigg)\delta E-\delta J,
\end{equation}

Substituting equations (\ref{z46}) and (\ref{z47}) into equation (\ref{z48}) yields
\begin{widetext}
\begin{equation}\label{z49}
\kappa^{'}M^{'2}-J^{'}=(M^2\kappa-J)+\bigg(2M\kappa+\frac{\beta}{M\kappa}\bigg)m^2\bigg(\frac{\omega}{m}-\frac{1}{2\kappa M+\frac{\beta}{M\kappa}}\bigg)\bigg(\frac{\omega}{m}-\Omega_h\bigg)[r_h^2+a^2]dt.
\end{equation}
\end{widetext}

1$\star$ In the extreme case, that is, when $M^2\kappa=J$, equation (\ref{z49}) then becomes
\begin{widetext}
\begin{equation}\label{z50}
\kappa^{'}M^{'2}-J^{'}=\bigg(2M\kappa+\frac{\beta}{M\kappa}\bigg)m^2\bigg(\frac{\omega}{m}-\frac{1}{2\kappa M+\frac{\beta}{M\kappa}}\bigg)\bigg(\frac{\omega}{m}-\Omega_h\bigg)[r_h^2+a^2]dt.
\end{equation}
\end{widetext}

Then, the angular velocity $\Omega_h$ can be expressed as
\begin{equation}\label{z51}
\Omega_h=\frac{a}{r_h^2+a^2}=\frac{M\kappa}{2M^2\kappa^2+\beta},
\end{equation}

When incident on the scalar field in the following mode
\begin{equation}
\frac{\omega}{m}=\frac{1}{2}\bigg(\frac{1}{2\kappa M+\frac{\beta}{M\kappa}}+\Omega_h\bigg),
\end{equation}

then equation (\ref{z50}) becomes
\begin{widetext}
\begin{equation}\label{z52}
\kappa^{'}M^{'2}-J^{'}=-\frac{1}{4}\bigg(2M\kappa+\frac{\beta}{M\kappa}\bigg)m^2\bigg(\Omega_h-\frac{1}{2\kappa M+\frac{\beta}{M\kappa}}\bigg)^2(r_h^2+a^2)dt,
\end{equation}
\end{widetext}

Combining equations (\ref{z51}) and (\ref{z52}), we obtain
\begin{equation}\label{z53}
\kappa^{'}M^{'2}-J^{'}=0.
\end{equation}

From equation (\ref{z53}), combined with previous analysis, this indicates that in the case of first order perturbations, TEH of a rotating short haired black hole cannot be disrupted in extreme conditions. Therefore, higher order cases will be considered next. In extreme conditions, when $\kappa M^2=J$, then equation (\ref{z45}) can be written as
\begin{equation}
\kappa^{'}M^{'2}-J^{'}=\bigg(\kappa+\frac{2\beta}{\kappa M^2}-\frac{\beta^2}{2M^4\kappa^3}-\frac{3\beta}{2\kappa M^2}\bigg)\delta E+o\delta E^3,
\end{equation}
which, through combining with $\beta=M^2-M^2\kappa^2$, leads to the arrangement of the above as
\begin{equation}\label{z54}
\kappa^{'}M^{'2}-J^{'}=\bigg(\frac{3}{2\kappa}-\frac{1}{2\kappa^3}\bigg)\delta E^2+o\delta E^3.
\end{equation}

In equation (\ref{z54}), there is this equation $\bigg(\frac{3}{2\kappa}-\frac{1}{2\kappa^3}\bigg)$, which is rewritten as $\frac{1}{2\kappa}\bigg(3-\frac{1}{\kappa^2}\bigg)$. Here, for $\kappa=\sqrt{1-\frac{\beta}{M^2}}$ to be meaningful, then $\kappa$ must be positive. Therefore, we only need to discuss the case of $\gamma_1=3-\frac{1}{\kappa^2}$ (see Figure (\ref{fig1})). One scenario is when $0<\kappa<\sqrt{\frac{1}{3}}$, leading to $\gamma_1=3-\frac{1}{\kappa^2}<0$, which results in $\kappa^{'}M^{'2}-J^{'}<0$. This indicates that in extreme conditions within this range, the incident scalar field in a rotating short haired black hole can disrupt the black hole's EH, meaning the weak cosmic censorship conjecture is violated. On the other hand, another scenario is when $\sqrt{\frac{1}{3}}<\kappa<1$, this range leads to $\kappa^{'}M^{'2}-J^{'}>0$. In other words, in extreme conditions within this range, TEH of a rotating short haired black hole cannot be disrupted.

2$\star$ When a scalar field is incident in a rotating short haired black hole, another case is the near extreme scenario, i.e., $J\ne M^2\kappa$, we have the following equation
\begin{widetext}
\begin{equation}\label{z55}
\kappa^{'}M^{'2}-J^{'}=(\kappa M^2-J)+\bigg(2M\kappa+\frac{\beta}{M\kappa}\bigg)m^2\bigg(\frac{\omega}{m}-\frac{1}{2\kappa M+\frac{\beta}{M\kappa}}\bigg)\bigg(\frac{\omega}{m}-\Omega_h\bigg)(r_h^2+a^2)dt,
\end{equation}
\end{widetext}

The mode of the incident scalar field in the near extreme scenario is
\begin{equation}
\frac{\omega}{m}=\frac{1}{2}\bigg(\frac{1}{2\kappa M+\frac{\beta}{M\kappa}}+\Omega_h\bigg),
\end{equation}

therefore, equation (\ref{z55}) can be expressed in the following manner.
\begin{widetext}
\begin{equation}\label{z58}
\kappa^{'}M^{'2}-J^{'}=(\kappa M^2-J)-\frac{1}{4}\frac{1}{\bigg(2M\kappa+\frac{\beta}{M\kappa}\bigg)}m^2\Omega_h^2\bigg(\frac{1}{\Omega_h}-\bigg(2\kappa M+\frac{\beta}{M\kappa}\bigg)\bigg)^2(r_h^2+a^2)dt,
\end{equation}
\end{widetext}

Here, the dimensionless infinitesimal parameter $\epsilon$, defined by equation (\ref{z25}), is used to represent the degree of approach
\begin{equation}
\frac{a^2+\beta}{M^2}=1-\epsilon^2,
\end{equation}

where $\epsilon$ is a small quantity approaching 0. Using Taylor expansion, we can obtain
\begin{equation}
a=\frac{J}{M}=M\kappa \bigg(1-\frac{\epsilon^2}{2\kappa^2}+o(\epsilon^4)\bigg),
\end{equation}

which is
\begin{equation}
J=M^2\kappa \bigg(1-\frac{\epsilon^2}{2\kappa^2}+o(\epsilon^4)\bigg),
\end{equation}

after calculation, we can obtain
\begin{equation}\label{z56}
M^2\kappa-J=\frac{M^2\epsilon^2}{2\kappa}-o(\epsilon^4),
\end{equation}

\begin{equation}\label{z57}
\frac{1}{\Omega_h}-2M\kappa-\frac{\beta}{\kappa M}=\frac{\bigg(2M^2\epsilon+\bigg(M^2+\frac{\beta}{2\kappa^2}\bigg)\epsilon^2-o(\epsilon^4)\bigg)}{a},
\end{equation}

substituting (\ref{z56}) and (\ref{z57}) into (\ref{z58}) yields the equation in the near extreme scenario
\begin{widetext}
\begin{equation}\label{z60}
\kappa^{'}M^{'2}-J^{'}=\bigg(\frac{M^2\epsilon^2}{2\kappa}-o(\epsilon^4)\bigg)-\frac{1}{4}\frac{1}{(\kappa M+\frac{M}{\kappa})}m^2\Omega_h^2\frac{\bigg(2M^2\epsilon+\bigg(M^2+\frac{\beta}{2\kappa^2}\bigg)\epsilon^2-o(\epsilon^4)\bigg)^2}{a}(r_h^2+a^2)dt.
\end{equation}
\end{widetext}

In Equation (\ref{z60}), we know that both $\epsilon$ and $dt$ are first order small quantities, the first parenthesis contains a second order small quantity, and the term following the first parenthesis is a third order small quantity. Through the analysis above, it is known that when $\kappa$ is positive, thus the value inside the first parenthesis is a positive second order small quantity, and the term following is a positive third order small quantity. Subtracting a positive third order small quantity from a positive second order small quantity results in an overall positive number. Therefore,
\begin{equation}\label{z61}
\kappa^{'}M^{'2}-J^{'}>0,
\end{equation}
 
Equation (\ref{z61})is derived, indicating that when a scalar field is incident on a rotating short hair black hole in near extreme conditions, TEH of this spacetime cannot be destroyed, and the singularity of the rotating short hair black hole is not exposed.

\section{summary}

In this paper, we utilize the exact solution of rotating short hair black holes to explore the weak cosmic censorship conjecture.

The results are as follows:

$\bullet$By introducing test particles into rotating short hair black holes, we discuss whether TEH of such black holes can be destroyed in the scenarios of extreme and near extreme black holes. We find that in the case of extreme black holes, considering the second order situation, when parameters meet the $0<\kappa<\sqrt{\frac{1}{3}}$ range, TEH of the rotating short hair black hole will be destroyed, exposing the singularity. Within the $\sqrt{\frac{1}{3}}<\kappa<1$ range, TEH of this spacetime cannot be destroyed. Another scenario is the near extreme case, where, under the second order approximation, within the $0<\kappa<1$ range, TEH of the rotating short hair black hole can be destroyed. Therefore, introducing test particles within the respective ranges will lead to a violation of the weak cosmic censorship conjecture.

$\bullet$By introducing a scalar field into rotating short hair black holes, we investigate whether TEH of such black holes can be destroyed in scenarios of extreme and near extreme black holes. We find that when the scalar field is introduced in extreme conditions, and $\kappa$ meets the $0<\kappa<\sqrt{\frac{1}{3}}$ range, TEH of the rotating short hair black hole is destroyed. In the $\sqrt{\frac{1}{3}}<\kappa<1$ range, TEH of the rotating short hair black hole cannot be destroyed. However, in near extreme conditions, TEH of the rotating short hair black hole cannot be destroyed.

In summary, if we impose numerical restrictions on the parameter $\kappa$ to ensure it falls within the corresponding range, this will lead to the violation of the weak cosmic censorship conjecture in rotating short hair black holes. On the other hand, there are scenarios where the weak cosmic censorship conjecture cannot be violated, which indirectly indicates that the conjecture adapts well to the context of rotating short hair black holes. This discovery lays the groundwork for our future research into the properties of rotating short hair black holes.

\begin{acknowledgments}
We acknowledge the anonymous referee for a constructive report that has significantly improved this paper. This work was supported by the Special Natural Science Fund of Guizhou University (Grants No. X2020068, No. X2022133) and the National Natural Science Foundation of China (Grant No. 12365008).
\end{acknowledgments}

\bibliographystyle{unsrt}
\bibliography{e2ref}

\end{document}